\documentstyle[aps,twocolumn,epsf]{revtex}
\begin{document}
\draft
\title{Destabilization of the 2D conducting phase by an in-plane magnetic 
field}
\author{J.S. Thakur and D. Neilson}
\address{School of Physics, The University of New South Wales, Sydney 2052 
Australia
\\[3pt]
\  \\ \medskip}\author{\small\parbox{14cm}{\small
We propose a mechanism for the recently reported destabilization by an
in-plane magnetic field of the conducting phase of low density
electrons in 2D.  We apply our self-consistent approach based on the
memory function formalism to the fully spin polarized electron system.
This takes into account both disorder and exchange-correlation
effects.  We show that spin polarization significantly favors
localization because of the enhancement of the exchange-correlations.
A key outcome is that the conducting phase for the fully spin polarized
system is significantly suppressed.  The in-plane magnetic field needed
to stabilize the fully spin polarized state lies in the range $0.1<H<1$
T, depending on the carrier density.  We determine the metal-insulator
phase diagram for the unpolarized and fully polarized systems, and we
estimate the dependence of the critical magnetic field on carrier
density.
 \\[3pt]{PACS numbers:
73.20.Dx,73.40.Qv,71.30.+h,71.55.-i}
}}\address{} \maketitle

\narrowtext

Even though the existence of a metal-insulator transition for
two-dimensional electron systems \cite{Kravchenko} has been known for
several years, the nature of the insulating and metallic states near
the transition is still a puzzle. In the presence of a magnetic field
which is perpendicular to the electron plane, the familiar quantum Hall
states are recovered \cite{Pudalov}. This is due to the dominant
contribution of orbital effects in the magneto-conductance. If the
magnetic field is parallel to the electron plane it can only couple
directly to the electron spin. Recent experiments
\cite{Simonian,Pudalov1,Hamilton} have reported that a weak parallel
magnetic field is sufficient to destroy the conducting phase making the
system insulating. The critical magnetic field needed varies with the
carrier density but is less than or of the order of $1$ T for both Si
and GaAs.

Numerical simulations of the interacting electron system in the absence
of imperfections in the substrate \cite{Senatore} have shown that for
electron densities $r_{s}\lesssim 20$, the ground state of the system
is the unpolarized electron liquid. As the electron density is lowered
and the electron correlations become stronger, the free energy per
electron for the fully spin polarized state approaches the free energy
of the unpolarized system. For $r_{s}\gg 10$ the free energies are very
close and the Zeeman energy gain from a quite small parallel magnetic
field could be sufficient to produce a fully polarized ground state.
The critical magnetic field needed to induce this transition will
become weaker for increasing $r_{s}$.

We have previously proposed that strong correlations in the presence of
weak disorder in the substrate can localize the electrons into a glassy
state \cite{TN1}. We have obtained reasonable agreement with the
position of the metal-insulator transition in zero magnetic field for
unpolarized electrons \cite{TN2}.  We know from numerical simulations
that spin polarized electrons are more strongly correlated than
unpolarized electrons at the same density \cite{Ceperley}. This is due
to the additional exchange contribution when all the electrons have
parallel spin. This suggests that at the same electron density and the
same level of disorder in the substrate, the polarized state is more
likely to be in an insulating state than the unpolarized state.

In our approach we determine the metal-glass transition using the
Kubo-relaxation function $\Phi _{\nu }(q,t)\equiv \langle N_{\nu
}(q,t)|N_{\nu }(q,0)\rangle .$ This is defined for the normalized
dynamical density variable $N_{\nu }(q,t)=\rho (q,t)/\sqrt{\chi _{\nu
}(q)}$, where $ \rho (q)=\sum_{k}a_{k}^{\dagger }a_{k+q}$ is the
density fluctuation operator. When the polarization index $\nu =p$ the
system is fully polarized with all the carrier spins aligned, while
$\nu =u$ is for the unpolarized system. $\chi _{\nu }(q)$ is the static
susceptibility for the corresponding system.  We are interested in the
dynamics of relaxation processes as $t\rightarrow \infty $.  The order
parameters for the glassy states are given by the relaxation function
in this limit, $f_{\nu }(q)=\lim_{t\rightarrow \infty }\Phi _{\nu
}(q,t)$. When $f_{\nu }(q)$ is non-zero, spontaneous fluctuations do
not decay even at infinite time.

Within the Mori-Zwanzig formalism \cite{Mori} we calculate $\Phi _{\nu
}(q,z) $ in terms of the Memory Function $M_{\nu }(q,z)$, 
\begin{equation}
\Phi_{\nu }(q,z)=-\frac{1}{z-\Omega_{\nu }(q)/(z+M_{\nu }(q,z))}, 
\label{Phi(q,z)}
\end{equation}
where $\Omega _{\nu }(q)=q^{2}/(m^{*}\chi _{\nu }(q))$.
In the limit $z\rightarrow 0 $ the relaxation function is \cite {TN1}
\begin{equation}
\lim_{z\rightarrow 0 }-z\Phi _{\nu }(q,z)=
f_{\nu }(q)=\frac{1}{1+{\Omega _{\nu }(q)}/{M_{\nu }(q)}},  \label{z->0f(q)}
\end{equation}
where $M_{\nu }(q) =\lim_{z\rightarrow 0} -zM_{\nu }(q,z)$.
This is evaluated using mode-coupling theory
\cite{MC}.  We obtain
\begin{equation}
M_{\nu }(q)=M_{\nu }^{cc}(q)+M_{\nu }^{ic}(q).
  \label{M(q)}
\end{equation}
$M_{\nu }^{cc}(q)$ is the contribution from interactions between
the carriers. 
The effect of scattering off disorder in the substrate is contained in $
M_{\nu }^{ic}(q)$.  Taking for the interaction between carriers 
$V(q)=2\pi e^{2}/\epsilon q$, where $\epsilon $ is 
the substrate dielectric constant,
we finally get for $M_{\nu }^{cc}(q)$ the expression 
\begin{eqnarray}
M_{\nu }^{cc}(q) =\frac{1}{2m^{*}q^{2}}\sum_{q^{\prime }}[
V(q^{\prime })({\bf q}\cdot {\bf q}^{\prime })+V(|{\bf q}-{\bf q}^{\prime
}|)\nonumber \\
\times({\bf q}\cdot ({\bf q}-{\bf q}^{\prime }))] ^{2}  
 \chi _{\nu }(q^{\prime })\chi _{\nu }(|{\bf q}-{\bf q}^{\prime
}|)f_{\nu }(q^{\prime })f_{\nu }(|{\bf q}-{\bf q}^{\prime }|).
\label{Mcc(q)}
\end{eqnarray}

 \begin{figure}
 \epsfxsize=8.0cm
 \epsffile{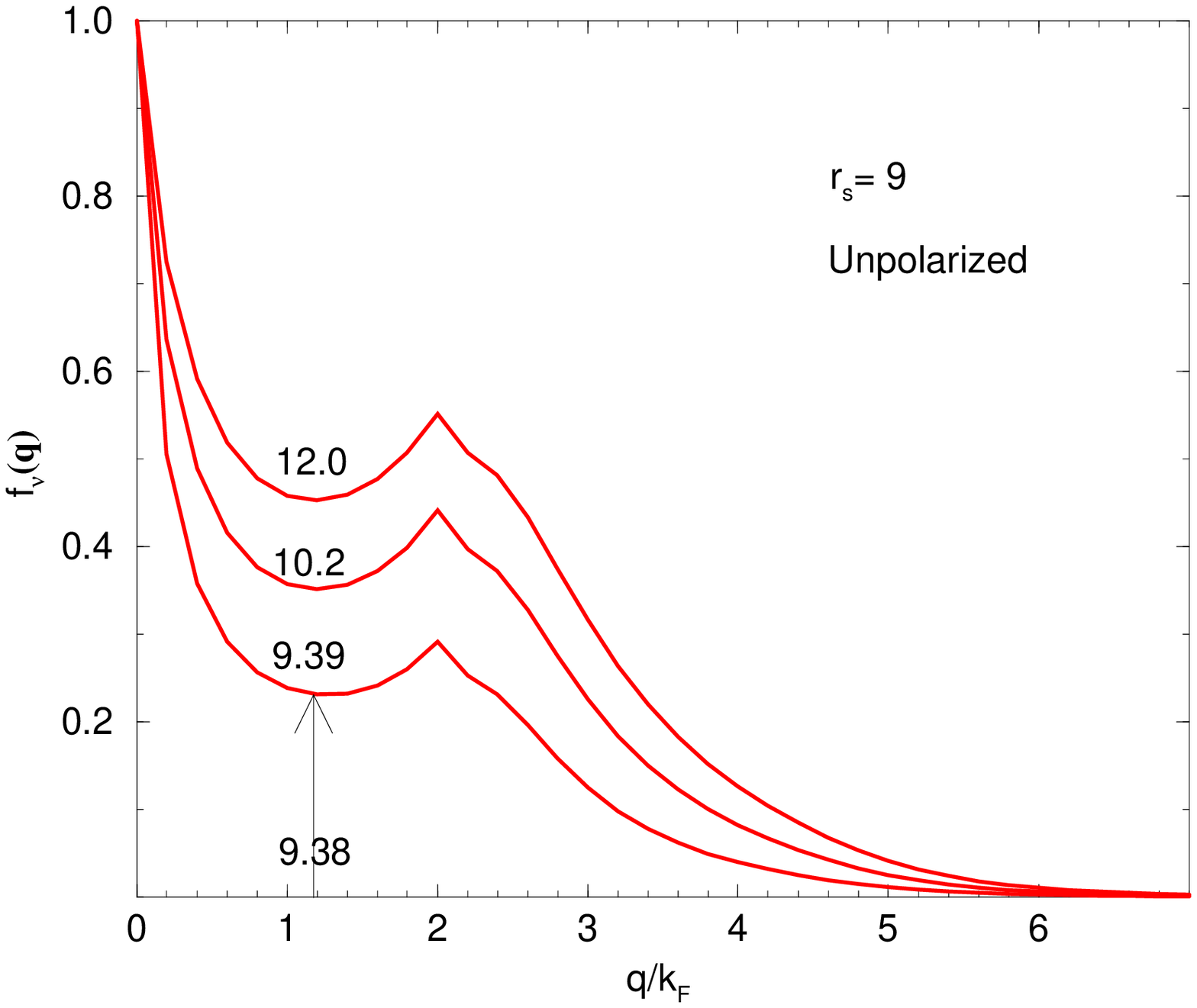}
\vspace{-1.0cm}
 \epsfxsize=8.0cm
 \epsffile{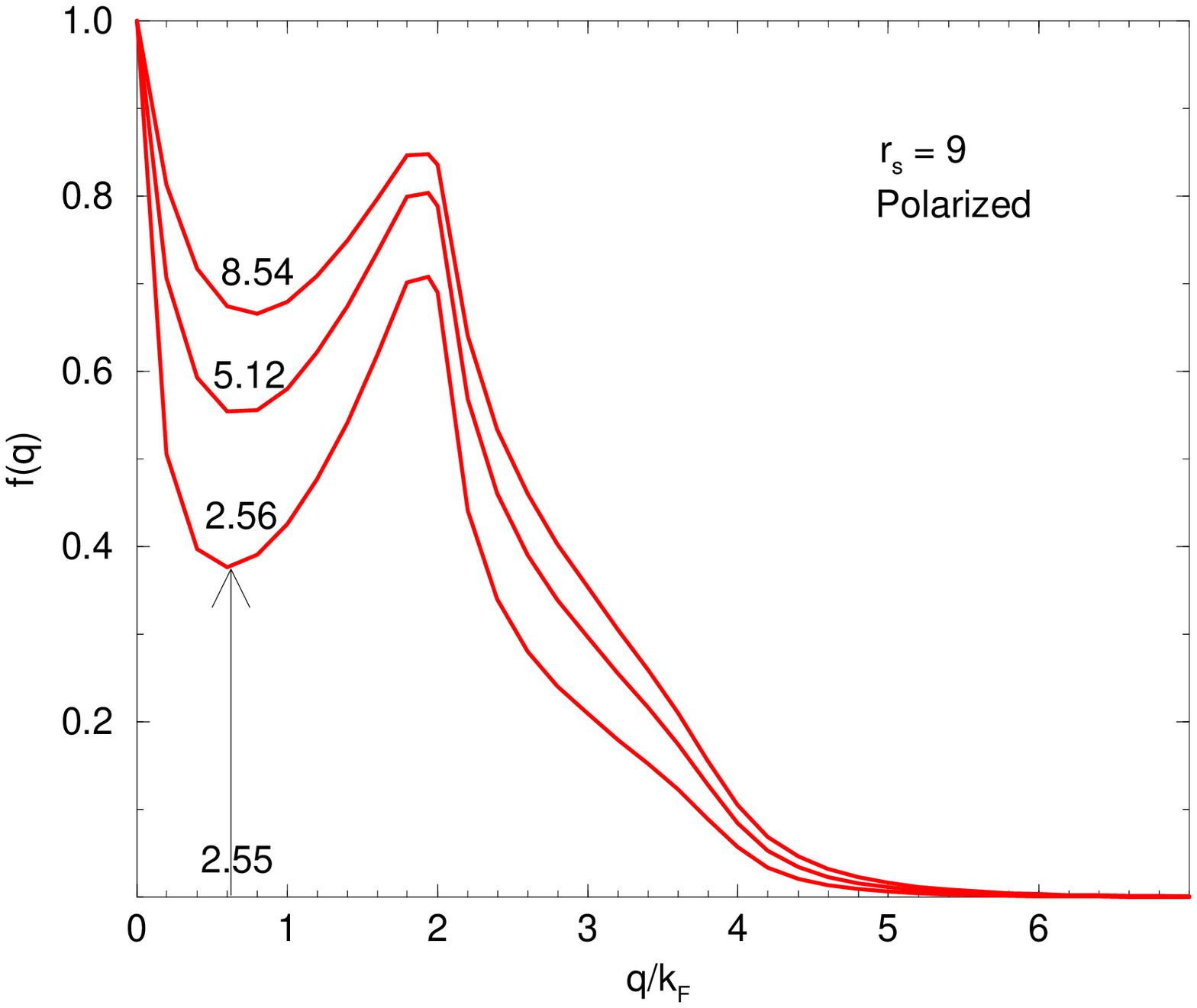}
  \caption[dummy3]{\widetext
Order parameters $f_{\nu }(q)$ for $r_{s}=9$ and $10$. Curve labels are
impurity densities $n_{i\text{ }}$ in units of $10^{9}$cm$^{-2}$.\\
 a. Unpolarized $r_s=9$.  $f_{u }(q)=0$ when
$n_{i\text{ }}<9.39\times10^{9} $cm$^{-2}$.
 b. Unpolarized $r_s=10$. $f_{u }(q)=0$ when
$n_{i\text{ }}<5.88\times10^{9} $cm$^{-2}$.\\ 
 c. Fully polarized $r_s=9$. $f_{p }(q)=0$ when
$n_{i\text{ }}<2.56\times10^{9} $cm$^{-2}$.
 d. Fully polarized $r_s=10$. $f_{p }(q)=0$ when
$n_{i\text{ }}<0.69\times10^{9} $cm$^{-2}$. 
 \label{f(q)}}
 \end{figure}
\narrowtext

For the disorder Memory Function $M_{\nu }^{ic}(q)$ 
we consider scattering off monovalent Coulomb impurities
randomly distributed within the plane of the carriers, $U_{\text{imp}
}(q)=[(2\pi e^{2})/(\epsilon q)]F_{i}(q)$, and scattering off the surface
roughness at the interface $W_{\text{surf}}(q)$. $F_{i}(q)$ is the impurity
form factor. We take $W_{\text{surf}}(q)=\sqrt{\pi }\Delta \Lambda \Gamma
(q) $exp$(-(q\Lambda )^{2}/8)$ appropriate for Si MOSFETs. Details of the
parameters used are given in \cite{TN1}. The final expression for 
$M_{\nu }^{ic}(q)$ is
\begin{eqnarray}
M_{\nu }^{ic}(q) &=&\frac{1}{m^{*}q^{2}}\sum_{q^{\prime }}\left[ n_{i}\langle
|U_{\text{imp}}(q)|^{2}\rangle +\langle |W_{\text{surf}}(q)|^{2}\rangle
\right]  \nonumber \\
&&\times ({\bf q}\cdot {\bf q}^{\prime })^{2}\chi _{\nu }(|{\bf q}-{\bf q}
^{\prime }|)f_{\nu }(|{\bf q}-{\bf q}^{\prime }|),
\label{Mic(q)}
\end{eqnarray}
where $n_{i}$ is the impurity density.  Equations \ref{z->0f(q)} to
\ref{Mic(q)} form a self-consistent set which we can solve to determine
the order parameters $f_{\nu }(q)$.

The correlations between carriers are taken into account through the
static susceptibilities $\chi _{\nu }(q)=\chi _{\nu }(q,\omega=0)$.
The $\chi _{\nu }(q)$ are known for the disorder free system from the
ground state properties of the system determined by numerical
simulations \cite{Ceperley}.  We write
 \begin{figure}
 \epsfxsize=8.0cm
 \epsffile{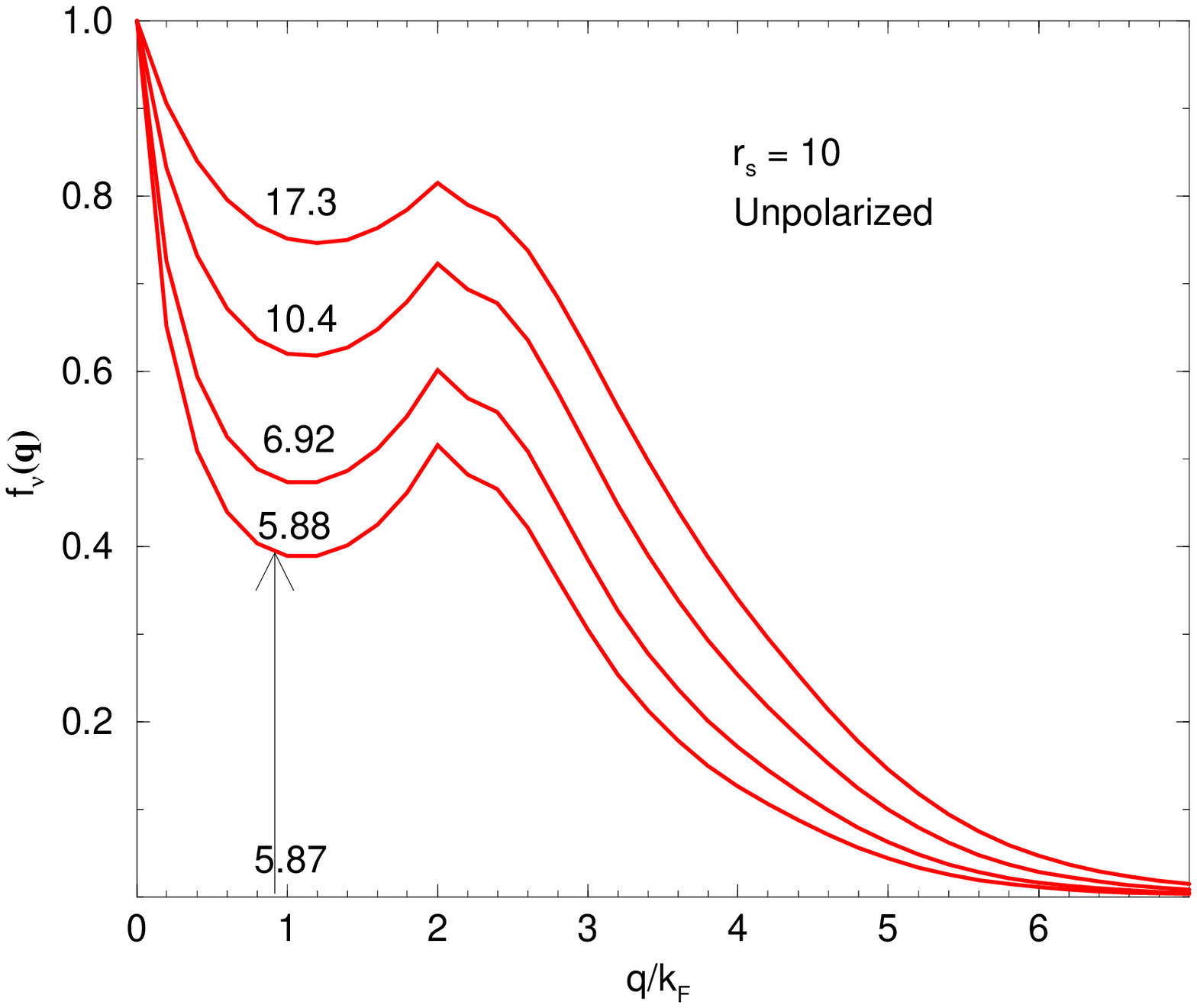}
\vspace{-1.0cm}
 \epsfxsize=8.0cm
 \epsffile{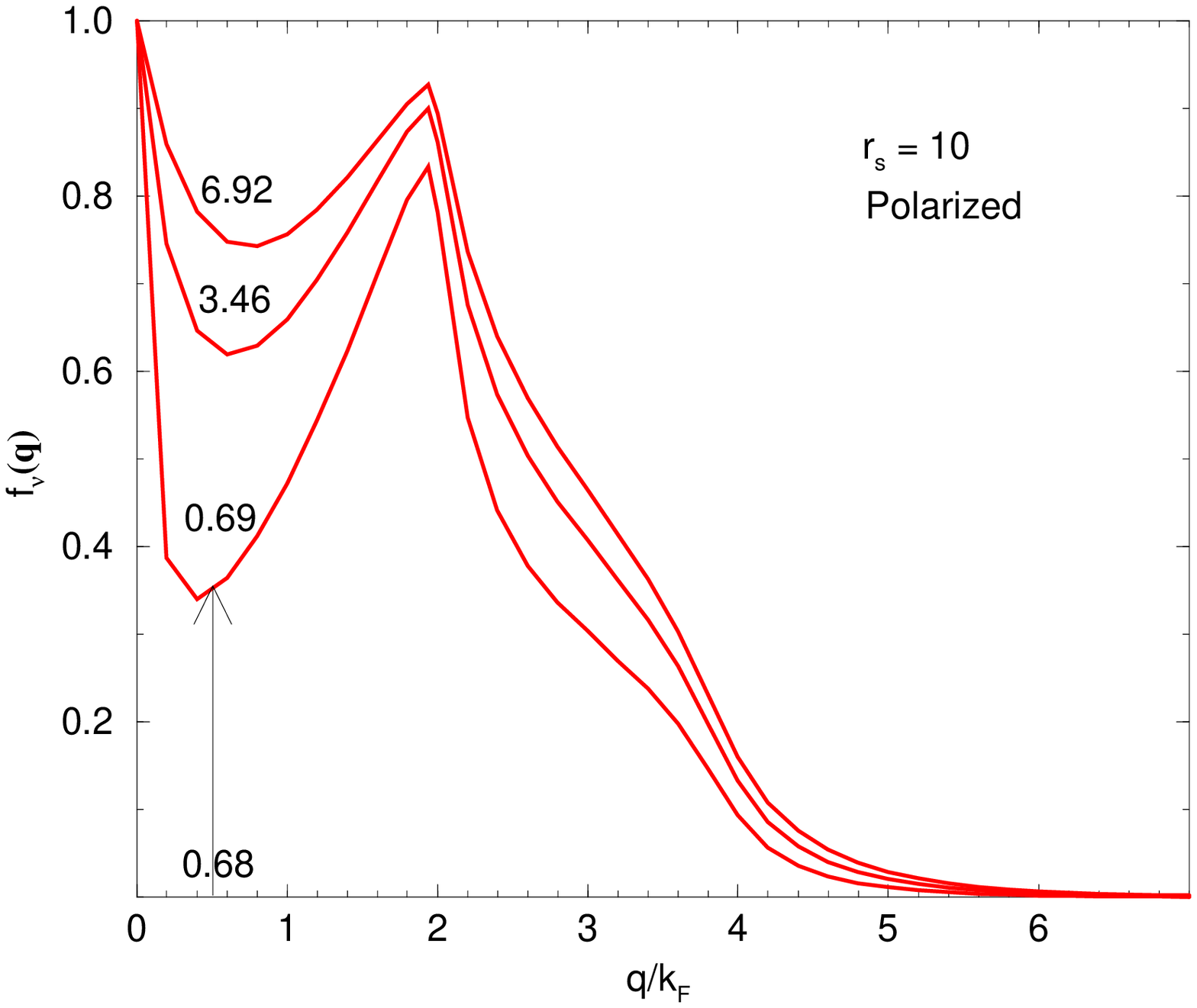}
 \end{figure}
\ \\ \ \\ 
\begin{equation}
\chi _{\nu }(q,\omega)=\frac{\chi _{\nu }^{(0)}(q,\omega)}
{1+V(q)[1-G_{\nu }(q)] \chi _{\nu }^{(0)}(q,\omega)}. 
 \label{chiv(q)}
\end{equation}
$\chi _{\nu }^{(0)}(q,\omega)$ is the Lindhard function. The static
local field factors $G_{\nu }(q)$ contain the correlations for the
polarization state $\nu$.  We use data from Ref.\ \cite{Ceperley} to
determine the $G_{\nu }(q)$ using Eq.\ \ref{chiv(q)} and the
fluctuation-dissipation theorem \cite{SNS}.

The level of disorder can be measured in terms of the scattering rate 
$\gamma _{\nu }$ calculated for carriers scattering from both impurities
of density $n_i$ and surface roughness.  To evaluate $\gamma _{\nu }$
we use the memory function formalism \cite{Gotze} to obtain,
\newpage
\begin{eqnarray}
{\rm i}\gamma _{\nu } &=&-\frac{1}{2m^{\star }n_{c}}\sum_{q}q^{2}\left[
n_{i}\langle |U_{\text{imp}}(q)|^{2}\rangle +\langle |W_{\text{surf}
}(q)|^{2}\rangle \right]  \nonumber \\
&&\times \left( \frac{\chi _{\nu }(q)}{\chi _{\nu }^{(0)}(q)}\right) ^{2}
\frac{\phi _{\nu }^{(0)}(q,{\rm i}\gamma _{\nu })}{1+{\rm i}\gamma \phi
_{\nu }^{(0)}(q,{\rm i}\gamma _{\nu })/\chi _{\nu }^{(0)}(q)}\ ,
\label{gamma}
\end{eqnarray}
where $\phi _{\nu }^{(0)}(q,{\rm i}\gamma _{\nu })=(1/{\rm i}\gamma
_{\nu })\left[ \chi _{\nu }^{(0)}(q,{\rm i}\gamma )-\chi _{\nu
}^{(0)}(q)\right] $ is the relaxation spectrum for non-interacting
carriers scattering off the disorder.  $\hbar\gamma _{\nu }$ is in
units of twice the Fermi energy.  From the scattering rate we calculate
the conductivity $\sigma$ at the transition using the Drude relation.

In Fig.\ 1 we show the order parameters $f_{\nu }(q)$ determined from
Eqs.\ \ref{z->0f(q)} to \ref{Mic(q)} for the polarized and unpolarized
states for a range of impurity densities $n_{i}$. When $n_{i}$ is less
than a critical density the $f_{\nu }(q)$ is zero, indicating a
conducting phase.  At the critical $n_{i}$ the $f_{\nu }(q)$ jumps
discontinuously, indicating a transition to an insulator \cite{TN1}.
The key point here is that for fixed $r_s$ the critical impurity
density is much smaller for the fully polarized system than it is for
the unpolarized system.

 \begin{figure}
 \epsfxsize=8.0cm
 \epsffile{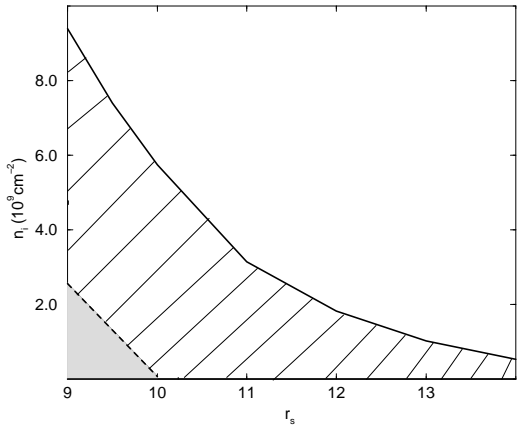}
  \caption[dummy3]{
Phase boundaries between conducting and insulating states for
unpolarized (solid line) and fully polarized systems (dashed line).
Axes are impurity density $n_i$ and $r_{s}$.  For fully polarized system
shaded region is conducting, and remaining area is insulating.   For
unpolarized system the conducting phase is the shaded plus hatched
regions, and remainder is insulating.
 \label{phasediagram}}
\end{figure}

We determined the critical $n_{i}$ for both the polarized and
unpolarized cases as a function of the carrier density.  Figure 2 shows
the resulting phase boundaries between conductor and insulator.  The
conducting phase for the fully polarized system which is represented by
the shaded region is restricted to a small range of $r_s$ below
$r_{s}\simeq10$.  In the absence of surface roughness scattering the
conducting phase extends to $r_s\simeq11$.  The conducting phase exists
only for small levels of disorder.  For the unpolarized system at the
same $r_{s}$ the critical level of disorder is significantly greater,
and the conducting phase extends to much larger values of $r_s$.  The
hatched region represents the reduction in the conducting phase region
when going from the unpolarized to fully polarized system.

Fig.\ 2 shows that fully spin polarizing the system destabilizes the
conducting phase except within a small range of carrier densities on
the higher density side. The stable conducting phase is restricted to
very small levels of disorder.  This significant shrinkage of the
conducting region is associated with the enhancement of
exchange-correlations for the fully polarized system.  This enhancement
favors localization.

We propose that the disappearance of the conducting phase in the
presence of an in-plane magnetic field is associated with polarization
of the carrier spins.  At these low carrier densities the energy cost
for spin aligned states becomes very small and a weak magnetic field is
sufficient to fully polarize the electrons.  Using numerical
simulations Rapisarda and Senatore \cite{Senatore} have calculated the
ground state energies $E_{p }$ and $E_{u }$ for the fully polarized and
unpolarized systems.  They report for $r_{s}\gtrsim 10$ the energies of
the two states differ by only a very small amount.

 \begin{figure}
 \epsfxsize=8.0cm
 \epsffile{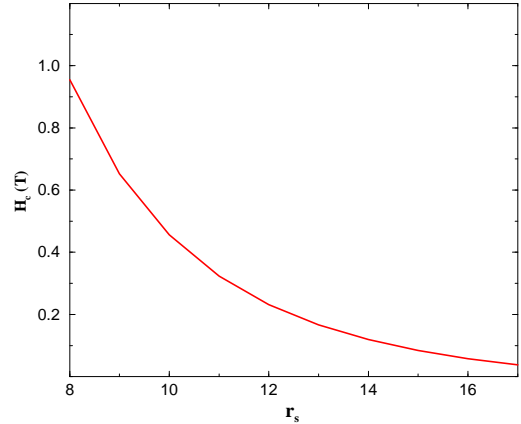}
  \caption[dummy3]{
Magnetic field at which Zeeman energy equals the energy difference
between the polarized and unpolarized states.
 \label{Hcvsrs}} \end{figure}

From these calculated $E_{p }$ and $E_{u }$ we can estimate the
critical magnetic field needed to drive the system into the fully spin
polarized ground state.  We equate the Zeeman energy splitting at the
critical field $H_{c}$ with the energy difference
\begin{equation}
\frac{g\mu _{B}}{\hbar }H_{c}=E_{p}-E_{u}. 
 \label{Hc}
\end{equation} 
In Fig.\ 3 we plot $H_{c}$ as a function of $r_{s}$ for holes in GaAs.
We use $(g\sigma_z)=1.1$ taken from Ref.\ \cite{Daneshvar}. The energies
$E_{p}$ and $E_{u}$ are calculated for the appropriate system
parameters.  

Hamilton {\it et al} \ \cite{Hamilton} reported for a GaAs sample with
hole density $p_s$ corresponding to $r_{s}=9$ that a magnetic field
$\alt0.7$ T drives the conducting state to an insulator. From Fig.\ 3
we find that at $r_{s}=9$ the critical magnetic field needed to fully
polarize the system is $H_{c}=0.6$ T, which is very close to this
value.  For electrons in Si MOSFETs the values of effective mass and
$(g\sigma_z)$ are not too different from those for holes in GaAs.  The
measured value of $H_{c}=0.5$ T in Si by Simonian {\it et al}
\cite{Simonian} at $r_{s}=9$ is also in good agreement with our value.

We find at $r_s=9$ that the critical disorder level needed to drive the
fully polarized system to the insulating state corresponds to a
conductivity of $\sigma\simeq4.5e^2/h$.  This is consistent with the
measured value at the transition of $\sigma\simeq5e^2/h$ for $r_s=9$
\cite{Hamilton}.

Hamilton {\it et al} \  give a phase diagram showing the
metal-insulator phase boundary as a function of $p_s$ and the magnetic
field.  For a conductivity $\sigma=5e^2/h$ we obtain an impurity
density for their sample of $n_i=2.4\times10^{9}$cm$^{-2}$.  Fig.\ 2
shows at this $n_i$ that the $r_s$ at the phase boundary drops from
$r_s=11.5$ in the unpolarized system ($H=0$) down to $r_s=9.2$ for the
fully polarized system ($H=0.6$~T).  To compare with the experimental
points taken from Ref.\ \cite{Hamilton}, we used a linear interpolation
between $p_s$ and $H$ and determined the corresponding critical
magnetic field as a function of $p_s$ (solid line).  Fig.\ 4 compares
the experimental points with our calculated $H_c$.  We find reasonable
agreement.  If we increase $n_i$ the solid line shifts to the right,
and correspondingly there is an increase in the critical magnetic field
(see Fig.\ 3).

 \begin{figure}
 \epsfxsize=8.0cm
 \epsffile{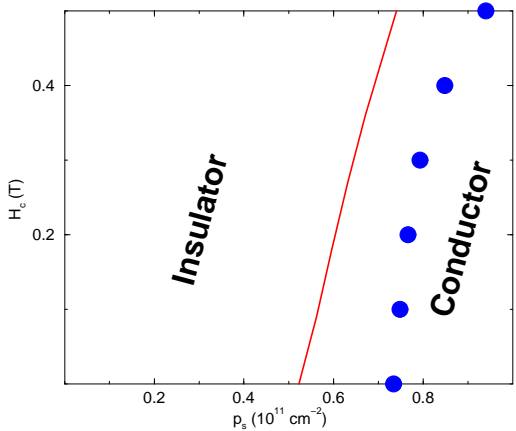}
  \caption[dummy3]{
Dependence of critical magnetic field for the metal-insulator
transition in GaAs on hole density $p_s$ (solid line) for impurity
density $n_i=2.4\times10^{9}$cm$^{-2}$.  Points are experimental data
taken from Hamilton {\it et al} \ \cite{Hamilton}.
 \label{Hcvsps}}
 \end{figure}

In conclusion we have demonstrated that magnetic fields $\alt1$ T
should be sufficient to fully spin polarize the carriers for $r_s>8$.
We have shown that the enhanced exchange-correlations for the fully
polarized system significantly favors the insulating phase.  Our
mechanism leads to results which are in reasonable quantitative
agreement with experiment.  We predict a re-emergence of a conducting
phase for the fully polarized system at very small levels of disorder,
but only for carrier densities $r_s<11$.

\acknowledgements

This work is supported by Australian Research Council Grant.
We thank Michelle Simmons and Leszek \'{S}wierkowski for very useful comments.


\begin{references}
\bibitem{Kravchenko} S.V. Kravchenko, G.V. Kravchenko, and J.E.
Furneaux, \prb {\bf 50}, 8039 (1994); S.V. Kravchenko, Whitney E.
Mason, G.E. Bowker, J. E. Furneaux, V.M. Pudalov, and  M. D'Iorio, \prb
{\bf 51} 7038 (1995); S.V. Kravchenko, D. Simonian, M.P. Sarachik,
Whitney Mason, and J.E. Furneaux, \prl {\bf 77}, 4938 (1996); Dragana
Popovic, A.B. Fowler, and S. Washburn, \prl {\bf 79}, 1543 (1997); M.Y.
Simmons, A.R. Hamilton, M. Pepper, E.H. Linfield, P.D.  Rose, D.A.
Ritchie, A.K. Savchenko and T.G. Griffiths \prl {\bf 80}, 1292 (1998);
M.Y. Simmons, A.R. Hamilton, T.G. Griffiths, A.K. Savchenko, M.
Pepper, and D.A. Ritchie, Physica B {\bf 251}, 705 (1998);  J. Lam, M.
D'Iorio, D. Brown and H. Lafontaine, \prb {\bf 56}, R12741 (1997); P.
T. Coleridge, R. L. Williams, Y. Feng, and P.  Zawadzki, \prb {\bf 56},
R12764 (1997)

\bibitem{Pudalov}  V.M. Pudalov, p.34 in {\em Proc. Int. Conf. on Electron
Localization and Quantum Transport in Solids}, Jaszowiec, Poland 1996, ed.
T. Dietl, Institute of Physics PAN, Warsaw (1996)

\bibitem{Simonian}D. Simonian, S.V. Kravchenko, M.P. Sarachik and V.M.
Pudalov, \prl {\bf 79}, 2304 (1997); 

\bibitem{Pudalov1}M. Pudalov, G. Brunthaler, A.  Prinz and G. Bauer,
Pis'ma Zh. Eksp. Teor. Fiz. {\bf 65}, 887 (1997) (JETP Lett. {\bf 65},
932 (1997))

\bibitem{Hamilton}  A.R. Hamilton, M.Y. Simmons, M. Pepper, E.H. Linfield,
P.D. Rose and D.A. Ritchie, preprint cond-mat/9808108 (1998)

\bibitem{Senatore}  Francesco Rapisarda and Gaetano Senatore, Aust J. Phys. 
{\bf 49, }161 (1996)

\bibitem{TN1}  J. S. Thakur and D. Neilson, Phys. Rev. {\bf 54}, 7674 (1996)

\bibitem{TN2}  J. S. Thakur and D. Neilson, Phys. Rev. Rapid Comm. (to
appear) (preprint cond-mat/9810288 (1998))

\bibitem{Ceperley}  B. Tanatar, and D. M. Ceperley, Phys. Rev. B {\bf 39},
5005 (1989)

\bibitem{Mori}  H. Mori, Prog.\ Theor.\ Phys.\ {\bf 33}, 423 (1965); R.
Zwanzig, in {\it Lectures in Theoretical Physics}, edited by W.E. Brittin,
B.W. Downs and J. Downs, Vol. {\bf 3} (Interscience, New York, 1961)

\bibitem{MC}U. Bengtzelious, W. G\"otze and A. Sj\"olander, J. Phys.
C. Solid State Phys. {\bf 17}, 5915 (1984); J. Bosse and J.S. Thakur,
Phys. Rev. Lett. {\bf 59}, 998 (1987)  

\bibitem{SNS}  L. \'{S}wierkowski, D. Neilson and J. Szym\'{a}nski, Phys.\
Rev.\ Lett.\ {\bf 67}, 240 (1991)

\bibitem{Gotze}  W. G\"{o}tze, Solid State Comm. {\bf 27}, 1393 (1978); A.
Gold and W. G\"{o}tze, {\it ibid.} {\bf 47}, 627 (1983)


\bibitem{Daneshvar}A.J. Daneshvar, C.J.B. Ford, M.Y. Simmons, A.V.
Khaetskii, A.R. Hamilton, M. Pepper and D.A. Ritchie, \prl {\bf 79},
4449 (1997)

 \end{references}
\end{document}